\documentclass[a4paper]{jpconf}
\pdfoutput=1

\usepackage{graphicx}
\usepackage{amsmath,amssymb,amsfonts}
\usepackage{subfigure}
\usepackage{listings}
\usepackage{cite}

\newcommand{\GOSAM}{{\textsc{Go\-Sam}}}
\newcommand{\GOLEMVC}{{\texttt{go\-lem95C}}}
\newcommand{\QGRAF}{{\texttt{QGRAF}}}
\newcommand{\FORM}{{\texttt{FORM}}}
\newcommand{\SPINNEY}{{\texttt{spin\-ney}}}
\newcommand{\HAGGIES}{{\texttt{hag\-gies}}}
\newcommand{\SAMURAI}{{\textsc{Sa\-mu\-rai}}}
\newcommand{\NINJA}{{\textsc{Nin\-ja}}}
\newcommand{\PYTHON}{{\texttt{Py\-thon}}}
\newcommand{\bea}{\begin{eqnarray}}
\newcommand{\eea}{\end{eqnarray}\noindent}

\newcommand{\bcen}{\begin{center}}
\newcommand{\ecen}{\end{center}}
\def\url#1{\texttt{#1}}

\begin{document}
\title{GoSam applications for automated NLO calculations}

\author{G.~Cullen}
\address{Deutsches Elektronen-Synchrotron DESY, Platanenallee 6, 15738 Zeuthen, Germany}

\author{H.~van Deurzen, N.~Greiner, G.~Heinrich\footnote{Speaker; talk given at the International Workshop on 
Advanced Computing and
Analysis Techniques in Physics Research (ACAT), Beijing, China, May 2013.}, G.~Luisoni, E.~Mirabella, 
T.~Peraro, J.~Reichel, J.~Schlenk, J.F. von Soden-Fraunhofen,}
\address{Max Planck Institute for Physics, F\"ohringer Ring 6, 80805 Munich, Germany}

\author{P.~Mastrolia}
\address{Max Planck Institute for Physics, Munich, Germany;\\
Dipartimento di Fisica e Astronomia, Universit\`a di Padova, and INFN Sezione di Padova, 
via Marzolo 8, 35131 Padova, Italy}

\author{G.~Ossola}
\address{Physics Department, New York City College of Technology, The
City University of New York,
300 Jay Street Brooklyn, NY 11201, USA;}
\address{The Graduate School and University Center, The City
University of New York,
365 Fifth Avenue, New York, NY 10016, USA}

\author{F.~Tramontano}
\address{Dipartimento di Scienze Fisiche, Universit\`a degli studi di Napoli and INFN, Sezione di Napoli, 
80125 Napoli, Italy}


\begin{abstract}
We present applications of the 
program \GOSAM{} for the automated calculation of 
one-loop amplitudes.
Results for  
NLO QCD corrections to beyond the Standard Model processes 
as well as Higgs plus up to three-jet production 
in gluon fusion are shown.
We also discuss some new features of the program.
\end{abstract}

\section{Introduction}

With the LHC data confirming the Standard Model 
to an almost incredible extent, precision measurements  
will be of enormous importance in order to 
scrutinize the Higgs properties and be sensitive to deviations from the 
Standard Model, and such measurements should come in hand with precision predictions.
Therefore, it is of primary interest to provide tools which allow one to perform
the comparison of LHC data to theory  at NLO accuracy.

The automation of NLO calculations for multi-particle final states has 
seen an enormous level of progress in the past years, as can be seen 
from examples for recent achievements in terms of 
codes\,\cite{Berger:2008sj,Bevilacqua:2011xh,Hirschi:2011pa,Cullen:2011ac,Cascioli:2011va,Agrawal:2012cv,Badger:2012pg,Actis:2012qn,Campbell:2012am,Arnold:2012xn} 
providing one-loop amplitudes.

In this talk, we explain the usage and present applications of the program \GOSAM{}\,\cite{Cullen:2011ac},  
which can generate and evaluate 
one-loop amplitudes for multi-particle processes in a fully automated way. 
\GOSAM{} also offers the possibility to be interfaced -- via 
the Binoth Les Houches interface\,\cite{Binoth:2010xt,Alioli:2013nda} -- to different 
Monte Carlo programs providing the real radiation and the infrared subtraction terms.
Further, an interface to model files in 
Universal FeynRules Output (\texttt{UFO})~\cite{Degrande:2011ua} or 
{\texttt LanHEP}~\cite{Semenov:2010qt} format 
allows to extend the application range of the code to Beyond the Standard Model physics.

\section{The \GOSAM{} program}

\subsection{Generation of the virtual amplitudes}

Amplitudes are expressed in terms of Feynman diagrams, 
where the integrand is generated analytically 
using \QGRAF~\cite{Nogueira:1991ex}, \FORM~\cite{Vermaseren:2000nd},
\SPINNEY~\cite{Cullen:2010jv} and \HAGGIES~\cite{Reiter:2009ts}. 

The information about the model
is either read from the built-in Standard Model file or
is generated from a 
Universal FeynRules Output (\texttt{UFO})~\cite{Christensen:2008py,Degrande:2011ua} or 
\texttt{LanHEP}~\cite{Semenov:2010qt} file.
Precompiled \texttt{MSSM\_UFO} and \texttt{MSSM\_LHEP} files 
and examples for their import
can also be found in the subdirectory \texttt{examples/model}.

The program offers the option to use different reduction techniques:
either the unitarity-based integrand reduction~\cite{Ossola:2006us,Ellis:2008ir}  
as implemented in \SAMURAI~\cite{Mastrolia:2010nb} 
or traditional tensor reduction as implemented in
{\tt golem95C}~\cite{Binoth:2008uq,Cullen:2011kv} interfaced through
tensorial reconstruction at the integrand level~\cite{Heinrich:2010ax},
or a combination of both.

\subsection{Installation and Usage}

The user can download the code  either as a tar-archive
or from the subversion repository at
    \bcen
    \url{http://projects.hepforge.org/gosam/}
    \ecen
To install \GOSAM{}, the user needs to run

\noindent {\tt python setup.py install --prefix MYPATH}\\
If \texttt{MYPATH} is not among the system default paths, 
the environment
variables \texttt{PATH}, \texttt{LD\_LIB\-RA\-RY\_PATH} and \texttt{PYTHONPATH}
might have to be set accordingly. For more details we direct the user to \cite{Cullen:2011ac}
and the reference manual coming with the code.
    
Prerequisites are a Linux/Unix environment, 
\PYTHON~($\geq2.6$), {\sc Java}~($\geq1.5$), {\tt Make}, and a {\tt Fortran95} compiler. 
On top of a standard Linux environment, the programs
\FORM\,\cite{Vermaseren:2000nd,Kuipers:2012rf} 
version~$\geq3.3$, and
\QGRAF\,\cite{Nogueira:1991ex} need to be installed on the system.
Further, at least one of the libraries
\SAMURAI{}\,\cite{Mastrolia:2010nb} or \GOLEMVC{}\,\cite{Cullen:2011kv}
needs to be present at compile time of the generated code.    
For the user's convenience we have prepared a package 
\texttt{gosam-\hspace{0pt}contrib-\hspace{0pt}1.0.tar.gz} containing
    \SAMURAI{} and \GOLEMVC{} together with the integral libraries
    \texttt{One\-LOop}\,\cite{vanHameren:2010cp},
    \texttt{QCD\-Loop}\,\cite{Ellis:2007qk} and
    \texttt{FF}\,\cite{vanOldenborgh:1989wn}.
   The package  is  available
   from \url{http://projects.hepforge.org/gosam/}.

\vspace*{3mm}

In order to generate the code for a process,
the user needs to prepare an input file 
which we will call \textit{process.in}, containing
\begin{itemize}
\item process specific information, such as a list of initial and
      final state particles, their helicities (optional),  
      the order of the coupling constants, and the underlying model;
\item scheme specific information, such as
      the regularisation and renormalisation schemes;
\item system specific information, such as paths to programs and libraries
      or compiler options;
\item optional information for optimisations which control the code generation.
\end{itemize}
The code can also generate a template input file.
In order to import settings with system specific information in an automated way,
the user can prepare a file {\it gosam.rc}  which will be imported into 
\textit{process.in} by
{\tt gosam.py -m gosam.rc -t process.in}.
The virtual amplitude can then be generated and compiled by invoking\\
{\tt gosam.py process.in}\\
{\tt make compile}




\subsection{Interfacing with Monte Carlo programs for the real radiation}
\label{sec:interface}

The so-called ``Binoth Les Houches Accord"~(BLHA)\,\cite{Binoth:2010xt,Alioli:2013nda}
defines an interface for a standardised communication between
one-loop programs (OLP) and Monte Carlo (MC) tools, where the latter 
provide the Born amplitude, as well as the matrix elements for the 
NLO real radiation and the infrared subtraction terms.
\GOSAM{} can act as an OLP in the framework of the BLHA, such that 
the calculation of complete cross sections is straightforward.

In the BLHA setup, the MC writes an order file containing the process specifications, 
called for  example 
\texttt{olp\_order.lh}, which can be used by \texttt{gosam.py}
to generate the virtual amplitude as follows:

\noindent {\tt gosam.py --olp --mc=MCname --config=YourPathTo/gosam.rc olp\_order.lh}

\noindent The sequence {\tt --mc=MCname} is optional, but can facilitate to adapt to MC specific settings.
Supported names at the moment are {\tt sherpa} and {\tt powhegbox}.
If {\tt gosam.rc} is in the current working directory or in the \GOSAM{} installation directory, 
its specification in the command line can be omitted.
The following sequence of commands will generate and compile the files for the virtual matrix
element:

\noindent
{\tt sh ./autogen.sh --prefix = `pwd'}\\
{\tt make install} \\
For more detailed information we refer to the {\tt BLHA HowTo} on the \GOSAM{} webpage.
Examples for full NLO calculations with \GOSAM{} interfaced to different 
MC programs are shown in Table~\ref{tab:interface}.
\begin{table}
\begin{center}
\begin{tabular}{|l|}
\hline
\hline\\
\GOSAM{} + MadDipole/MadGraph4/MadEvent\,\cite{Frederix:2008hu,Frederix:2010cj,Gehrmann:2010ry,Stelzer:1994ta,Maltoni:2002qb,Alwall:2007st}\\
\hline\\
$pp\to b\bar{b} b\bar{b}$\,\cite{Binoth:2009rv,Greiner:2011mp}\\
$pp\to W^+W^-$+2\,jets\,\cite{Greiner:2012im}\\
$pp\to \gamma\gamma$+1,2\,jets \,\cite{Gehrmann:2013aga,Gehrmann:2013bga} \\
SUSY QCD corrections to $pp\to \tilde{\chi}_1^0 \tilde{\chi}_1^0$+jet \,\cite{Cullen:2012eh} \\
QCD corrections to $pp\to $ graviton($\to \gamma\gamma$)+jet \,\cite{Greiner:2013gca} \\
 \hline
 \hline\\
\GOSAM{} + {\sc Sherpa}\,\cite{Gleisberg:2007md,Gleisberg:2008ta}\\
\hline
$pp\to W^+W^+ +$2\,jets\\
$pp\to W^\pm b\bar{b}$\\
$pp\to W^+W^-\,b\bar{b}$\,\cite{wwbb}\\
$pp\to W^+W^-$\\
$pp\to W^\pm$+0,1,2,3 jets\\
$pp\to Z/\gamma$+0,1,2 jets\\
$pp\to t\bar{t}$+0,1 jets\,\cite{Hoeche:2013mua}\\
$pp\to H+$2\,jets (gluon fusion)\,\cite{vanDeurzen:2013rv}\\
$pp\to t\bar{t}H$+0,1 jet\,\cite{vanDeurzen:2013xla} \\
 \hline
 \hline\\
\GOSAM{} + MadDipole/MadGraph4/MadEvent+{\sc Sherpa}\\
\hline
$pp\to H+$3\,jets (gluon fusion) \, \cite{Cullen:2013saa}\\
\hline
 \hline\\
\GOSAM{} +{\sc Powheg}\,\cite{Frixione:2007vw,Alioli:2010xd}\\
\hline
$pp\to HW/HZ$+0 and 1 jet\,\cite{Luisoni:2013cuh}\\
\quad\\
\hline
\end{tabular}
\end{center}
\caption{NLO calculations done by interfacing \GOSAM{} with different Monte Carlo programs.   
\label{tab:interface}}
\end{table}
Further, 
pre-generated code to be used within {\sc Sherpa} for a large number of processes can be 
downloaded from
\url{http://projects.hepforge.org/gosam/proc}.
To produce results for these processes, no  
\GOSAM{} installation is needed. Automated scripts 
coming with the process packages will ensure smooth running within the {\sc Sherpa} framework.

\section{Phenomenological results}

In the following we show a selection of results obtained with \GOSAM{} interfaced to different Monte Carlo programs. 



\subsection{$W^+W^-\,b\bar{b}$}

We  calculated the NLO QCD corrections to the  process 
$pp (p\bar{p}) \to W^+W^-b\bar{b}+X \to (e^+ \nu_e)(\mu^-
\bar{\nu}_{\mu})\,b\bar{b}+X$,  
leading to a final state which 
is a signature of the decay of a $t\bar{t}$ pair with leptonic $W$ boson decays, 
including singly-resonant and non-resonant contributions.
Results are shown for the LHC at $7$\,TeV. 
All final state partons are clustered into jets with a separation $R>0.5$ 
using the anti-$k_T$ jet algorithm \cite{Cacciari:2005hq,Cacciari:2008gp} implemented 
in {\tt Fastjet} \cite{Cacciari:2011ma}. 
Each event has to contain at least two $b$-jets with
  $ p_{T,b} >30$\,GeV and $\eta_{b} < 2.5$.
Further cuts are    $p_{T,l} >20\,\text{GeV},   \eta_{l} < 2.5, 
   p_{T,\text{miss}} >20$\,GeV.
\begin{figure}[htb]
\subfigure[]{\includegraphics[width=7.cm]{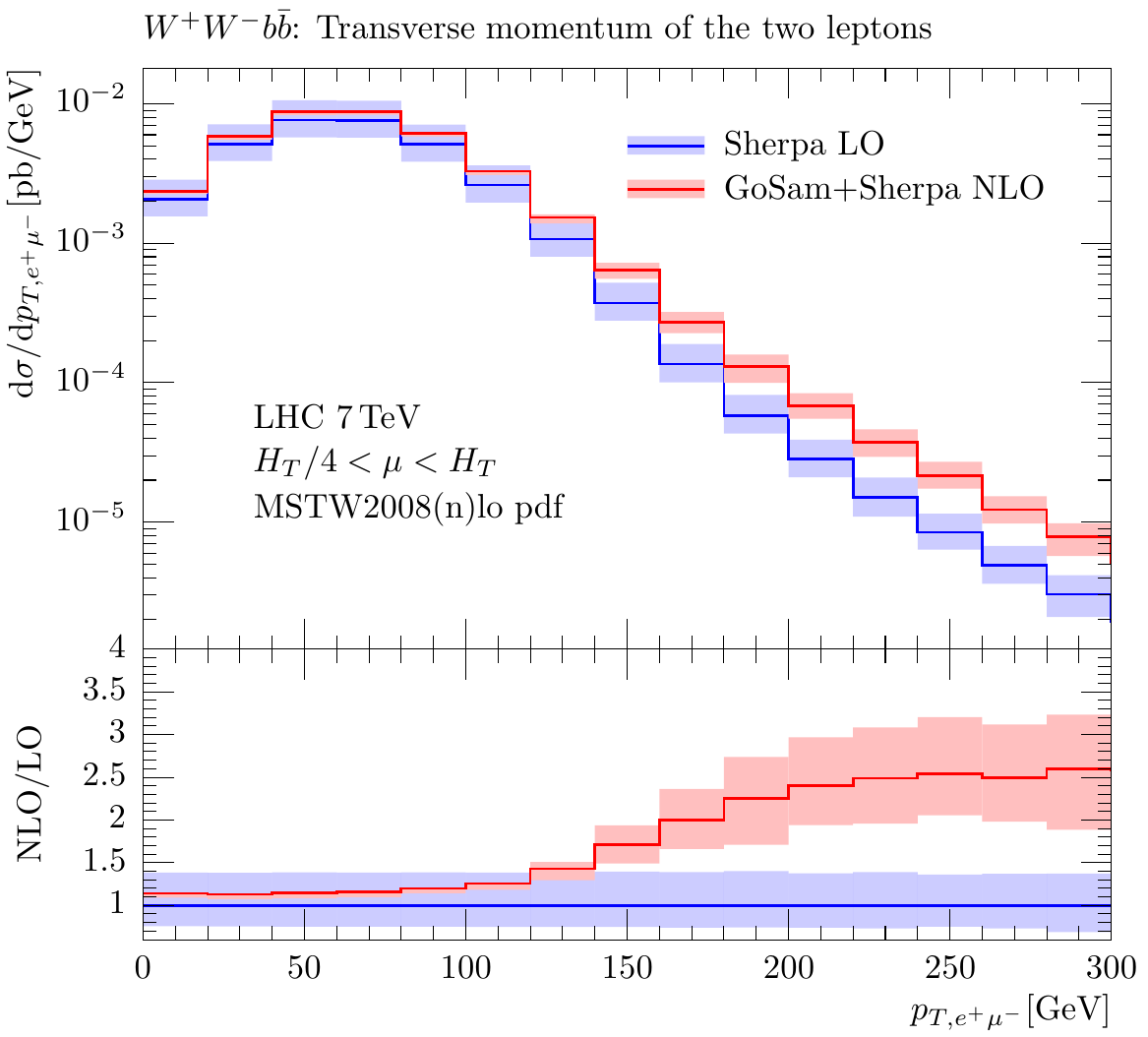} }\hfill
\subfigure[]{\includegraphics[width=7.cm]{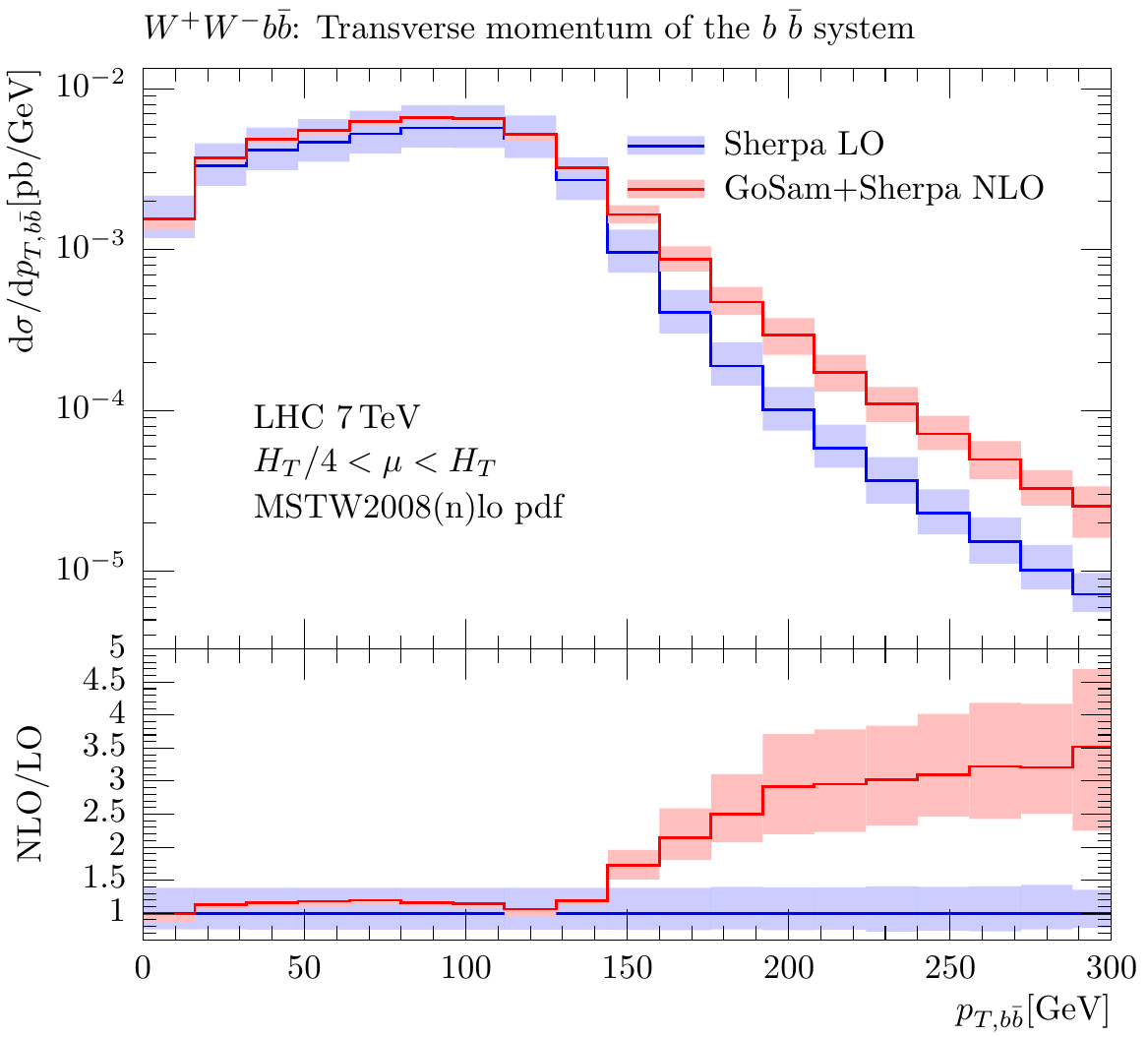} }
\caption{Results for (a) the transverse momentum distribution of the
  lepton pair stemming from $W^+W^-$ boson decay and (b) transverse
  momentum distribution of the sum of the two tagged $b$-jet momenta.}
\label{fig:wwbb}
\end{figure}

Figure \ref{fig:wwbb} shows the  distributions for the sum of the
transverse momenta of the two leptons and the two $b$-jets, respectively.  These
observables receive large NLO corrections because most of the
particles inherit their transverse momentum from a top quark pair. At
LO the $t\bar{t}$ pair has zero transverse momentum, while it can
obtain transverse momentum at NLO by recoiling against the real
radiation.  
Note that the LO scale variations cannot account for this effect and therefore  
the uncertainty band on the LO distribution obtained
by scale variations does not include the NLO result in the tail of
these distributions.

\subsection{Diphotons+jets}

\begin{figure}[htb]
\subfigure[]{\includegraphics[width=7.5cm]{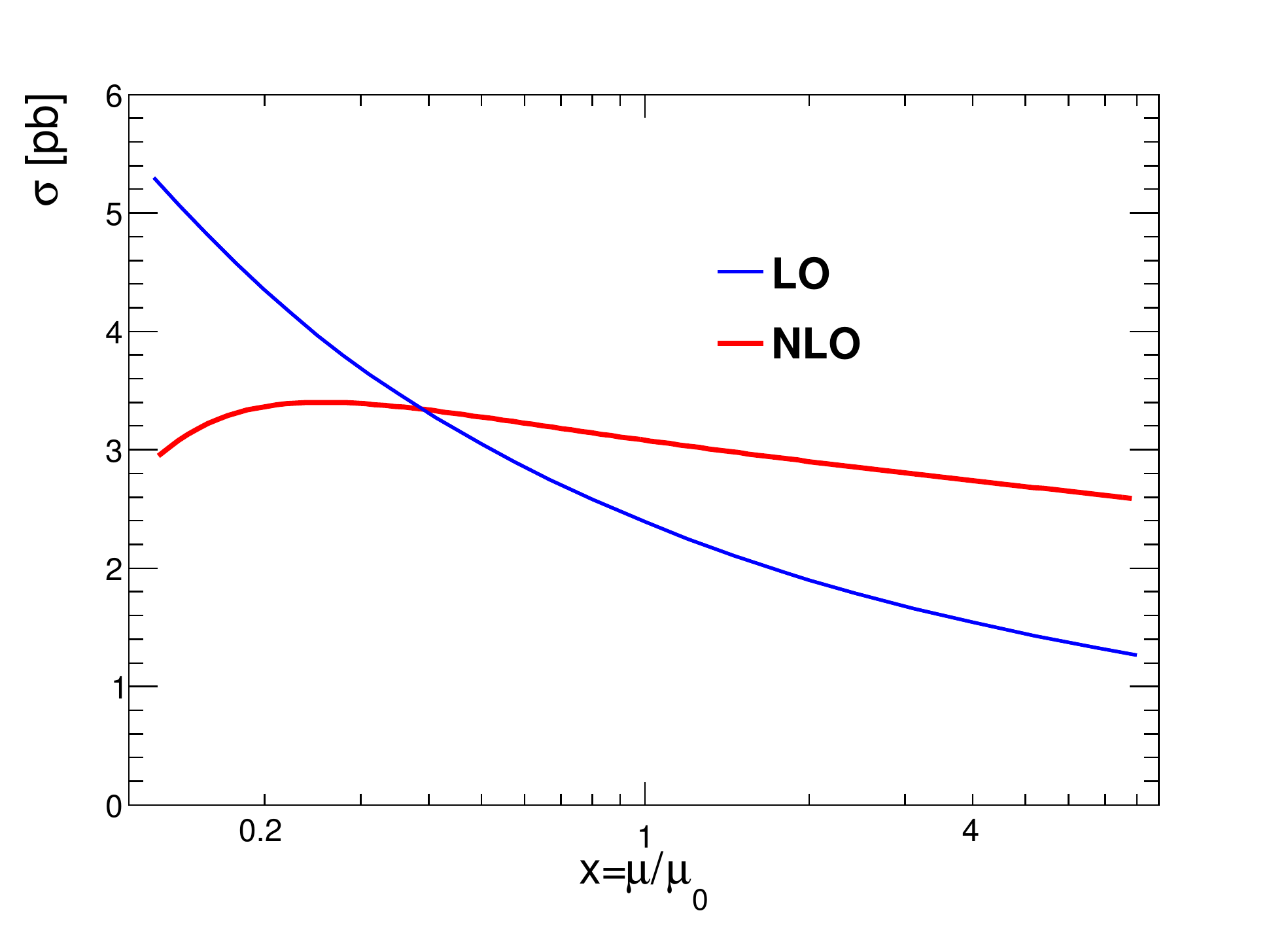} }\hfill
\subfigure[]{\includegraphics[width=7.cm]{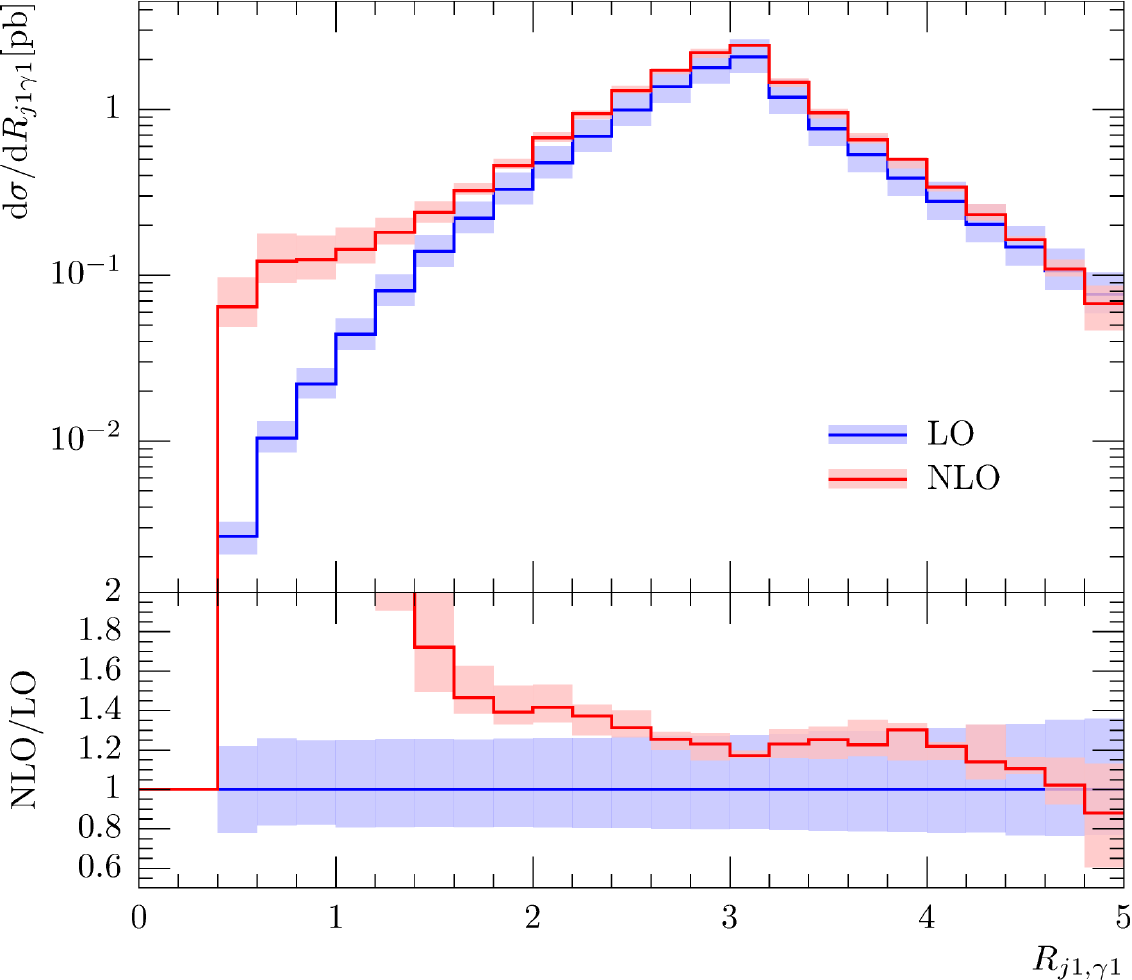} }
\caption{(a) scale variations for diphoton+2 jets, (b) $R$-separation between the hardest jet and the hardest photon.}
\label{fig:diphoton}
\end{figure}

\GOSAM{} in combination with MadDipole/MadGraph4/MadEvent\,\cite{Frederix:2008hu,Frederix:2010cj,Gehrmann:2010ry,Stelzer:1994ta,Maltoni:2002qb,Alwall:2007st} has also been used to calculate the NLO QCD corrections to 
$pp\to\gamma\gamma +1$\,jet\,\cite{Gehrmann:2013aga} and $pp\to\gamma\gamma +2$\,jets\,\cite{Gehrmann:2013bga},
where the former also includes the fragmentation component.
The numerical results for $pp\to\gamma\gamma +2$\,jets which are shown in Fig.~\ref{fig:diphoton} 
have been calculated at
a center-of-mass energy  $\sqrt{s}=8$\,TeV, using 
the smooth cone isolation
criterion~\cite{Frixione:1998hn} with 
$R=0.4, n=1$ and $\epsilon=0.05$. Renormalization and factorization scales $\mu$ and $\mu_F$
have been set equal, 
with the central scale $\mu_0^2= \frac{1}{4}\,(m_{\gamma\gamma}^2+\sum_j p_{T,j}^2)$.
For the  jet clustering we used an anti-$k_T$ algorithm~\cite{Cacciari:2008gp} with a cone size
of $R_j=0.5$ provided by 
the {\tt FastJet} package \cite{Cacciari:2011ma}.
The following kinematic cuts have been applied:
\begin{eqnarray*}
&&
p_T^{\rm{jet}}>30\mbox{ GeV},\quad  p_T^{\gamma,1}>40 \mbox{~GeV}, \quad
p_T^{\gamma,2}>25\mbox{~GeV},\\
&&|\eta^{\gamma}| \leq 2.5, \quad
|\eta^{j}| \leq 4.7,\quad R_{\gamma ,j} > 0.5,\quad R_{\gamma, \gamma} >0.45.
\end{eqnarray*}
Again, we observe a significant change of the shape at NLO, due to the extra QCD radiation in regions which are kinematically suppressed at LO.

\subsection{Associated Higgs production}

After the recent discovery of a Higgs boson at the LHC, being able to
disentangle signal processes from background ones and also the
different production channels of the SM Higgs boson became of central
relevance. The developments in \GOSAM{} allowed recently to compute
the NLO QCD corrections to the production of $H+2$
jets~\cite{vanDeurzen:2013rv} and $H+3$ jets~\cite{Cullen:2013saa} (in
gluon-gluon fusion in the $m_{top}\to \infty$ limit) and also of
$Ht\bar{t}$ and $Ht\bar{t}+$jet~\cite{vanDeurzen:2013xla} for the LHC
at 8 TeV. $H+2$ jets and the processes involving the top quarks were
computed using a fully automated interface to the Sherpa MC event
generator, based on the BLHA, whereas for $H+3$ jets a hybrid setup
combining the virtual part generated by \GOSAM{} with
MadDipole/Madgraph4/MadEvent and Sherpa was used. The virtual
corrections for $Ht\bar{t}(j)$ where computed using a new reduction
algorithm based on an integrand decomposition via Laurent expansion\,\cite{Mastrolia:2012bu},
which was implemented in the library \NINJA{}.

In the calculation of $H+3$ jets the cteq6L1 and cteq6mE
parton distribution functions were used for LO and NLO respectively,
and a minimal set of cuts based on the anti-$k_T$ jet algorithm with
$R=0.5$, $p_{T,min}=20$ GeV and $\left|\eta\right|<4.0$ was applied.
For $Ht\bar{t}(j)$ we used CT10 at NLO and a minumum transverse
momeutm cut of $p_{T,min}=15$ GeV.

\begin{figure}[htb]
\subfigure[]{\includegraphics[width=7.cm]{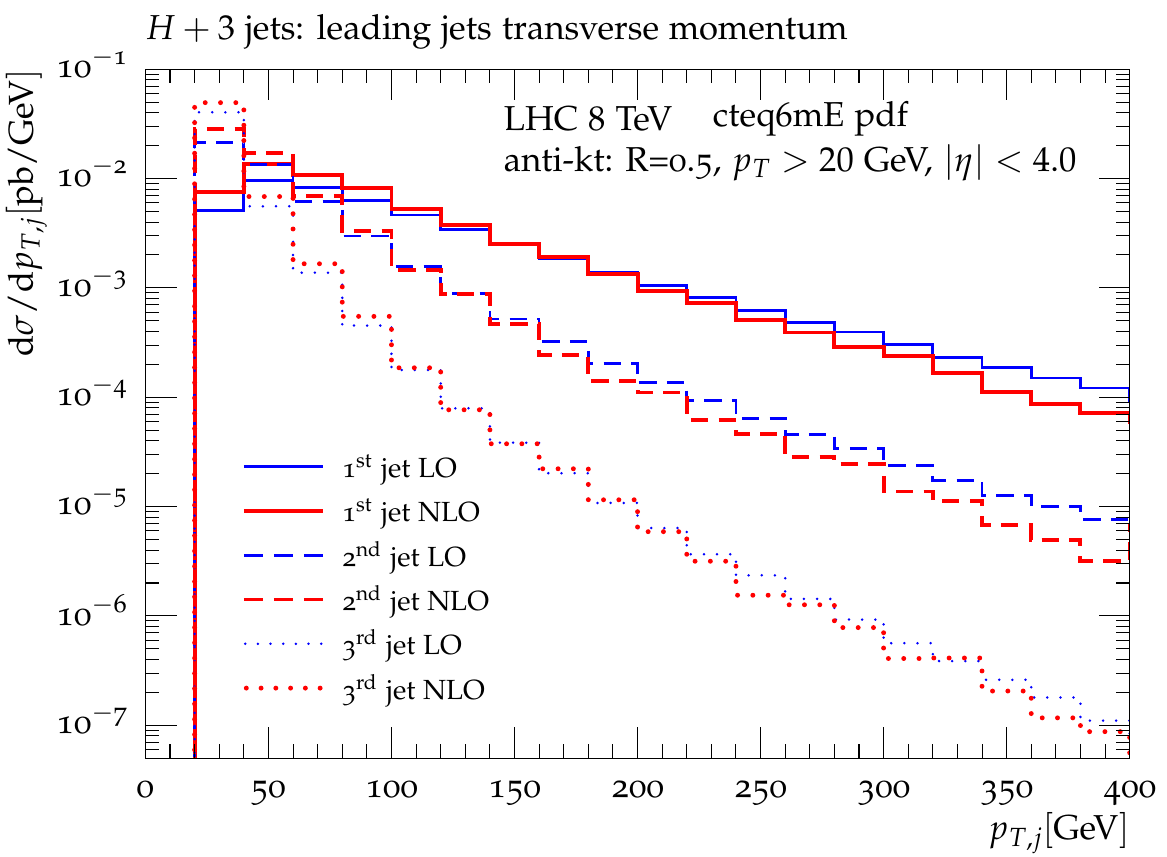} }\hfill
\subfigure[]{\includegraphics[width=7.cm]{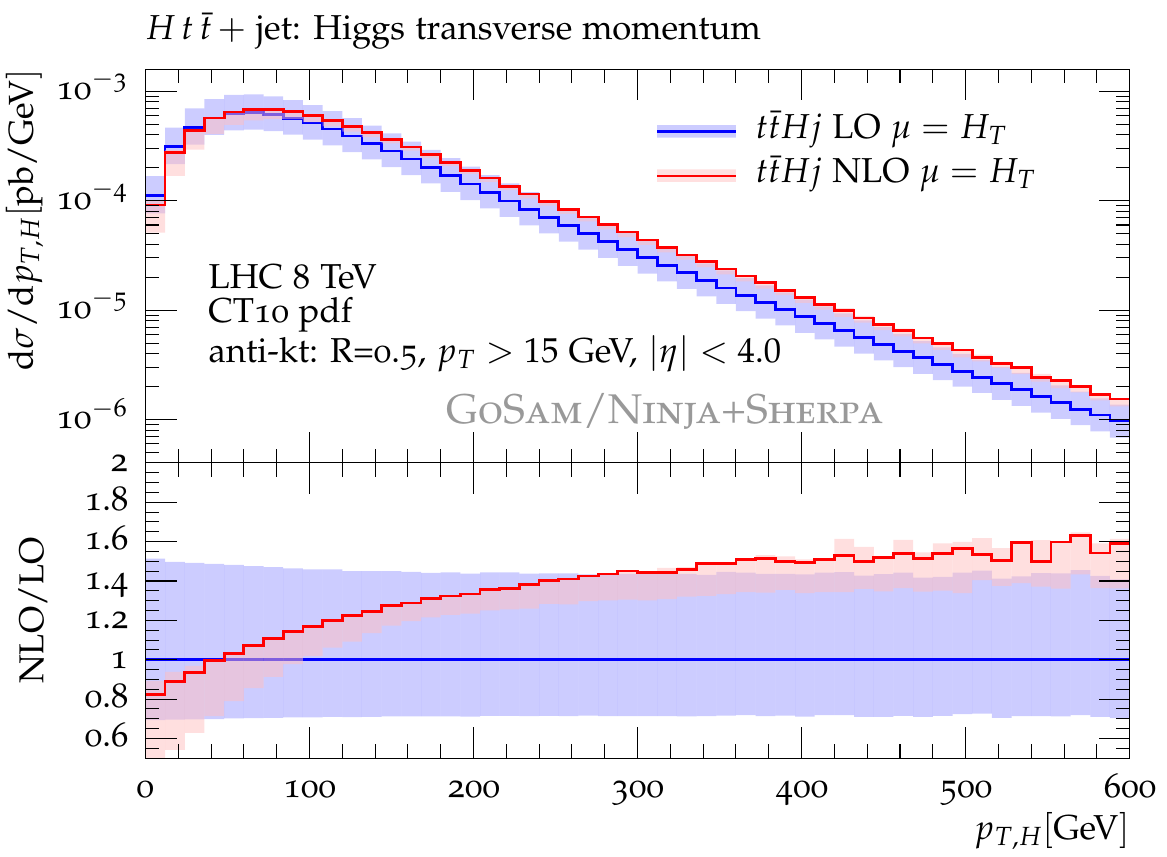}}
\caption{(a) transverse momentum distributions of the leading jets in
  $H+3$ jets production, (b) transverse momentum distribution of the
  Higgs boson in $Ht\bar{t}+$jet production.}
\label{fig:H3j}
\end{figure}

Figure~\ref{fig:H3j}(a) shows the transverse momentum distributions at
LO and NLO for the three leading jets in $H+3$ jets. We observe
that the NLO corrections drag the $p_T$-spectra towards smaller $p_T$ values, 
as expected as an effect of additional QCD radiation.
Figure~\ref{fig:H3j}(b) displays the transverse
momentum distribution of the Higgs boson in $Httj$ at LO and NLO. In
this case the NLO corrections become larger with increasing
$p_T$. Furthermore the scale uncertainty is reduced by 60-70\% in
going from LO to NLO.

\subsection{SUSY-QCD corrections to neutralino pair plus jet production}

\begin{figure}[htb]
\subfigure[]{\includegraphics[width=9.3cm]{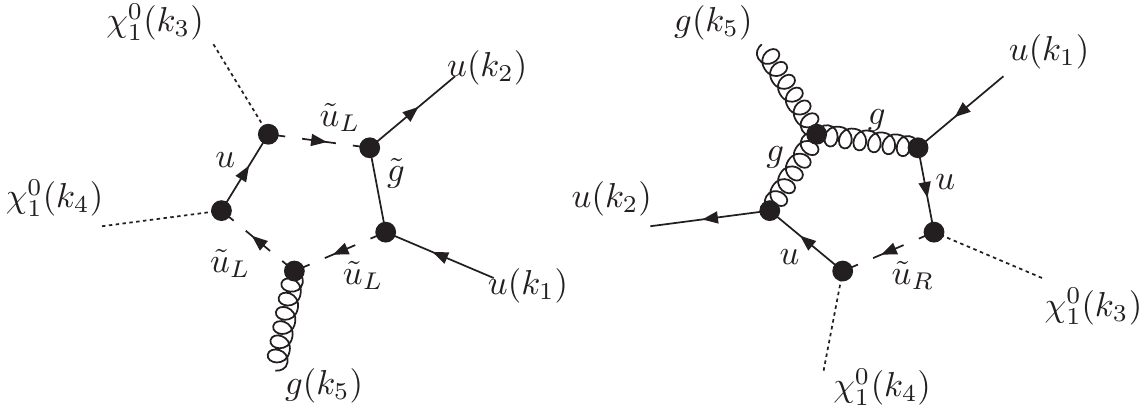} }\hfill
\subfigure[]{\includegraphics[width=7.cm]{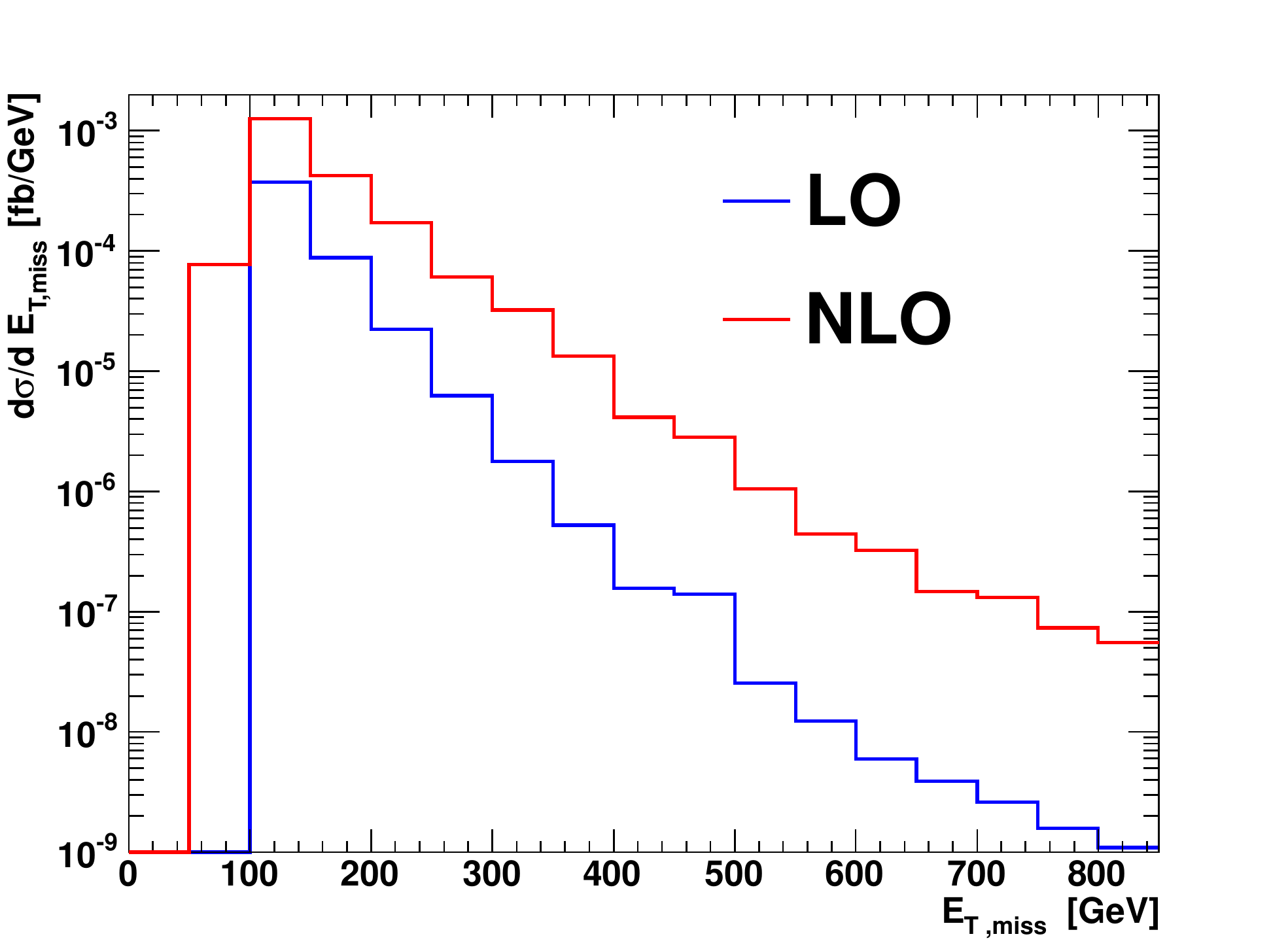} }
\caption{(a) Two illustrative pentagon diagrams, (b)  missing transverse energy $E_T^{\rm{miss}}$ for the process $pp\to \tilde{\chi}_1^0\tilde{\chi}_1^0$+jet at $\sqrt{s}=8$\,TeV.}
\label{fig:susy}
\end{figure}

\GOSAM{} also has been used to calculate the NLO Susy-QCD corrections to the production of 
a pair of the lightest neutralinos plus 
one jet at the LHC at $8$\,TeV, 
appearing as a monojet signature in combination with missing energy. 
We fully included all non-resonant diagrams, 
i.e. we did not use the simplifying assumption that production and decay
factorise. 
Examples of pentagon diagrams occurring in the virtual corrections, 
as well as the missing transverse energy distribution, are shown in Fig.~\ref{fig:susy}.
We observe that the NLO corrections are large, mainly due to additional channels opening up at NLO.
The detailed setup can be found in \cite{Cullen:2012eh}.

\subsection{\GOSAM{} and extra dimensions}

\begin{figure}[htb]
\subfigure[]{\includegraphics[width=7.cm]{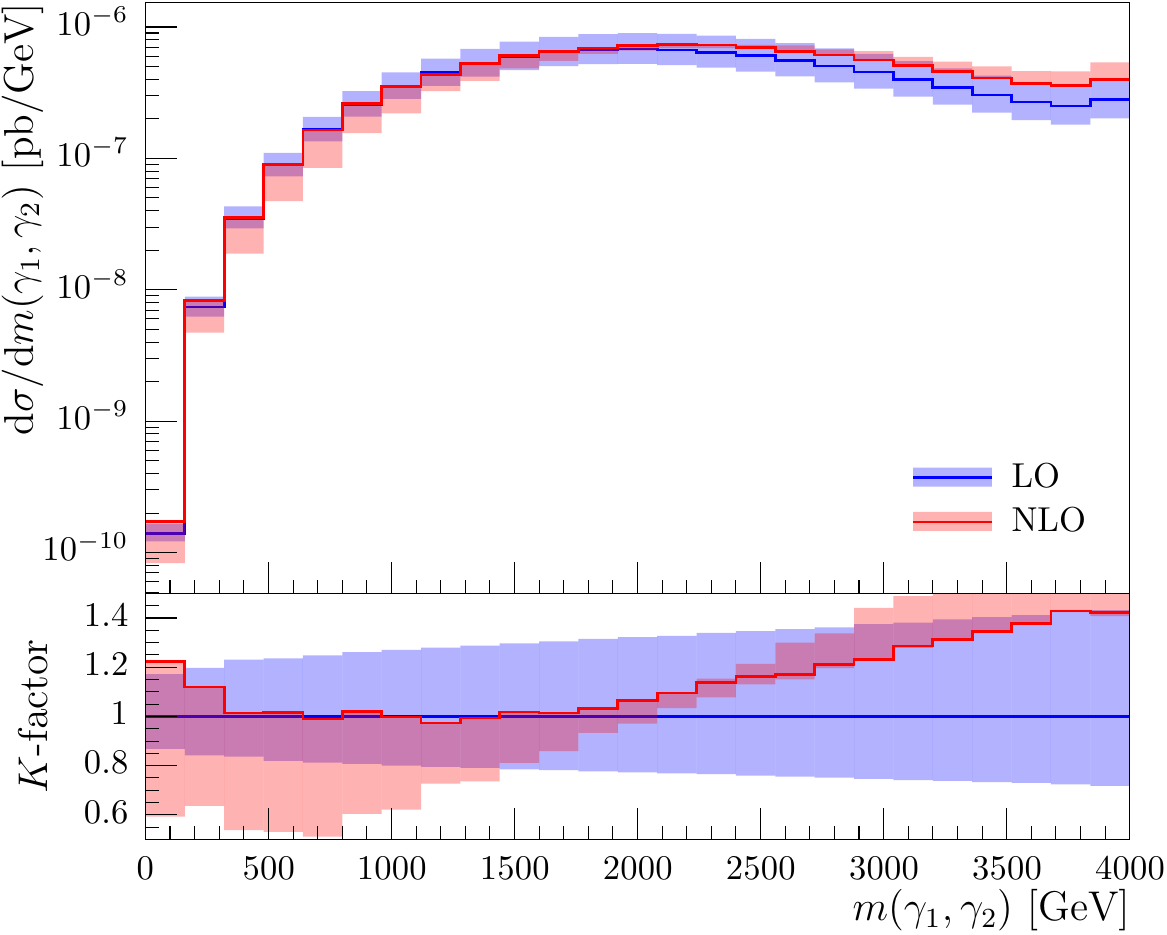} }\hfill
\subfigure[]{\includegraphics[width=7.cm]{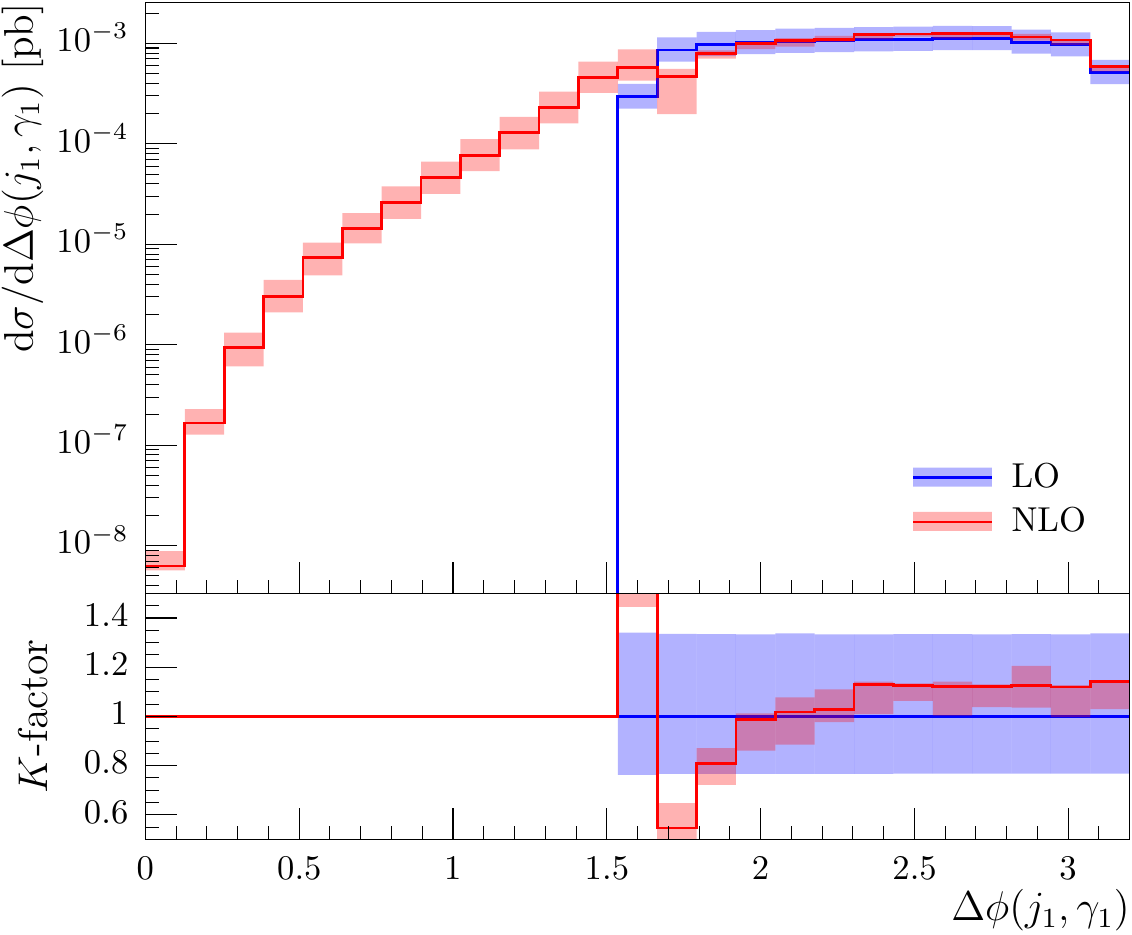} }
\caption{(a) NLO QCD corrections to the invariant mass distribution of the photon pair stemming from graviton decay
within the ADD model for $\delta=4$ large extra dimensions.  
The bands show the scale variations by a factor of two around the central scale 
$\mu_0^2 = \mu_F^2 = \frac{1}{4} \left(  m_{\gamma\gamma}^2 +  p_{T,jet}^2  \right)$.
(b) Azimuthal angle distribution between the leading photon and the leading jet.
The region $\Delta\phi<\pi/2$ is kinematically inaccessible at LO.}
\label{fig:extradim}
 \end{figure}

Another computationally intense calculation based on \GOSAM{}+
MadDipole/MadGraph4 
are the NLO QCD corrections to the production of 
a graviton in association with one jet\,\cite{Greiner:2013gca}, 
where the  graviton decays into  a photon pair,   
within ADD models of large extra dimensions\,\cite{ArkaniHamed:1998rs,Antoniadis:1998ig}.
The calculation is quite involved due to the complicated tensor structure 
introduced by spin-2 particles, and the non-standard propagator of the graviton, 
coming from the summation over the very densely distributed Kaluza-Klein modes.
As can be seen from Fig.~\ref{fig:extradim}, the 
K-factors turn out {\it not} to be 
uniform over the range of the diphoton invariant mass distribution. 
As the latter in general is used to derive exclusion limits, the 
differential NLO corrections
should be taken into account.
For details we refer to \cite{Greiner:2013gca}.

\section{Code development}

We are working on a number of new features concerning code generation 
as well as integrand reduction.
For instance, we implemented a 
 new strategy to produce optimized {\tt fortran95} code 
 based on   {\tt FORM} version 4\,\cite{Kuipers:2012rf}
 rather than {\tt haggies}\,\cite{Reiter:2009ts}, leading to
faster code generation and more compact code.
Further, the possibility of
parallelisation  at diagram level, the option to have   numerical polarisation vectors
and the option to sum diagrams sharing the same propagators algebraically at 
 {\tt FORM} level lead to an
enormous gain in code generation time and reduction of code size.
Concerning the amplitude reduction, we implemented 
integrals where the rank exceeds the number of propagators. 
An alternative reduction based on the Laurent expansion 
method developed in \cite{Mastrolia:2012bu}
also has been implemented and used successfully in \cite{vanDeurzen:2013xla}.

A version \GOSAM\,2.0 where all these new features are publicly available is in preparation.


\section{Conclusions}

We have presented applications of the program \GOSAM{} 
which can generate and evaluate one-loop matrix elements for multi-particle processes 
in an automated way.
The program is publicly available at
\url{http://projects.hepforge.org/gosam/} 
and can be used to produce NLO corrections
within QCD, electroweak theory, or other models 
which can be imported  via an interface to
FeynRules. 
Monte Carlo programs for the real radiation are linked 
via the BLHA (Binoth Les Houches Accord) interface.
This way \GOSAM{} is a very flexible and widely applicable tool 
for the automated calculation of multi-particle observables at next-to-leading order.

\subsection*{Acknowledgments}
The work of G.C. was supported by the DFG SFB-TR-9 and the
EU TMR Network LHCPhenoNet.
P.M. is supported by the Alexander von
Humboldt Foundation, in the framework of the Sofja Kovaleskaja Award Project
``Advanced Mathematical Methods for Particle Physics'', endowed by the German
Federal Ministry of Education and Research.
The work of G.O. was supported in part by the National Science Foundation
under Grant PHY-1068550.


\section*{References}


\providecommand{\newblock}{}

\end{document}